\documentstyle[11pt,newpasp,twoside,psfig]{article}
\begin{document}
\title{Proper Motion of Water Masers Near NGC1333-SVS13}
\author{Alwyn Wootten }
\affil{NRAO, 520 Edgemont Road, Charlottesville, VA 22903}
\author{Kevin Marvel}
\affil{American Astronomical Society, 2000 Florida Avenue, NW, Suite 400, Washington,
DC  20009}
\author{Mark Claussen}
\affil{NRAO,P.O.Box O,Socorro,NM,87801}
\author{Bruce Wilking}
\affil{U. Mo.-St Louis}

\begin{abstract}
VLBA observations of water masers toward the region of NGC1333 near SVS13, 
the driving source for the well-known Herbig-Haro objects HH7-11, are reported. 
Maser emission was observed over four epochs spaced by three weeks during 
late 1998.   SVS13 is associated with the millimeter continuum SED Class I 
source SSV13A1.  High resolution CO observations have secured the association 
of the flow with the continuum object on arc-second scales.  A 1998 Aug 12 
VLA observation places the blueshifted masers within 80 AU (0''.2) of the 
2.7 mm position for A1 reported from BIMA.  We report observations of two 
groups of masers, one redshifted by about 6 km s$^{-1}$ and one blueshifted 
by about  2-3 km s$^{-1}$, separated by about 100 AU in projected distance 
along position angle 21 degrees.  Redshifted masers were present only in 
the latter two epochs.  During all epochs, an arclike structure was present 
with similar morphology; over time this structure moved relative to the 
southernmost blue maser toward the southeast, roughly along the position 
angle defined by the Herbig-Haro objects, with a proper motion of about 
13 km s$^{-1}$.  Both redshifted and blueshifted masers are clearly part 
of the same lobe of the flow; their observed Doppler shifts are artifacts 
of the opening angle of the flow.  Our data suggest that the opening angle 
is about 10$^o$.  We measure lower expansion velocities and a wider opening 
angle than has been found in some other flows near Class 0 SED objects, 
perhaps indicating that the flows open and slow as they age.

\end{abstract}

\section{Introduction}
The tremendous brightness of water masers make them good tracers of motions in molecular gas on small scales.  The Very Long Baseline Array  (VLBA) provides beams of about one half milliarcsecond (0.17 AU) in size and can measure proper motions of about 10 km s$^{-1}$ in the several days to weeks lifetime of a typical maser at a distance of 350 pc, the distance generally employed for NGC1333.   Observationally, masers occur within about 100 AU of their low luminosity driving source, and provide excellent signposts for the source of a flow, as well as good probes of the flow geometry and dynamics. The geometry of a flow and the physics of maser emission (produced along a tangential sight-line through the flow cocoon) conspire to result in a spread of radial velocity in a typical low mass maser source of only a few  km s$^{-1}$.  A few Herbig-Haro objects have measured proper motions of as much as a few hundred  km s$^{-1}$; CO also shows very high velocities in this source--can the masers also show such high space velocities?  Here we report VLBA observations toward the region of NGC1333 near SSV13, the driving source for the well-known Herbig-Haro objects HH7-11.

\section{Observations and Discussion}

NGC1333 was observed with the 10 stations of the Very Long Baseline Array and a single antenna from the Very Large Array, both facilities of the National Radio Astronomy Observatory, on 1998 September 15, October 4, October 27 and November 14. Approximately four hours per epoch were devoted to observations of NGC1333.  Observations of strong continuum sources were made for calibration purposes.

The VLBA data were recorded with a 4 MHz bandwidth centered at a velocity of 7.0 km s$^{-1}$ relative to the local standard of rest (LSR).  Both right and left circular polarizations were recorded with 1-bit sampling, and correlated with the NRAO VLBA correlator to provide 256 spectral channels for each polarization averaged every 2 seconds.  This correlator mode provided a velocity resolution of 0.21 km s$^{-1}$ per channel and a maximum field of  view of 2.5\arcsec. 

	At millimeter wavelengths, SSV13 consists of at least three components, named A, B and C by Looney, Mundy and Welch (2000).  The combined emission from these comprises at least part of
the emission from IRAS 3 (Jennings et al. 1987).  SSV13 itself is the infrared counterpart of source A; the centimeter counterpart is called VLA4 ( Rodriguez, Anglada and Curiel 1997).  This is generally accepted as the powering source for the outflow associated with the chain of HH objects HH7-11.  Blueshifted molecular gas in this outflow is associated with HH7-11 to the southeast while redshifted gas lies to the northwest (Knee and Sandell 2000). Source A, in turn, appears to  have a weak companion, named A2 by Looney, Mundy and Welch, which corresponds to centimeter source VLA3 (Rodriguez, Anglada and Curiel 1997).  It was proposed as a possible powering source for the outflow originating in this region by those authors. Looney, Mundy and Welch note  that its diffuse structure, its weakness and absence in 1.3mm images  (Bachiller et al. 1998) suggest it to be dominated by free-free emission. Water maser emission was first remarked in the region near A by Haschick et al. (1980); the accuracy of their position is not sufficient to determine whether source A1 or A2 might be the origin. The maser emission reported in the present paper appears to be associated with A1. The optically invisible source to the southwest, C or MMS3, was a source of water maser emission and named H$_2$O-B by Haschick et al. (1980) but has not been active during our observations. The central source, SSV13B, corresponds to a weak centimeter wave source, VLA 17.  This source lacks an optical or near-infrared counterpart and powers a remarkably collimated SiO flow (Bachiller et al. 1998), with red gas to the north and blue gas to the south.

	Throughout the epoch of observation, three main groupings of blueshifted masers occurred. The brightest, Group I, lay to the SW, with Group II defining an arc structure  to the East and Group III dominated by a single bright spot at Epoch II and III to the North of Group I.  All masers were somewhat blueshifted with respect to the ambient velocity determined from SiO observations by Codella et al. (1999) of 8 km s$^{-1}$; those in the arc lie at velocities between 6 and 7 km s$^{-1}$ while the other two groups lie at about  5.5 km s$^{-1}$.  During the last two epochs, redshifted masers appeared near 14.1 km s$^{-1}$.  The axis defined by the normal to the arc structure of Group II, and which runs between Group I and Group III, lies at position angle $30^o \pm  5^o$ in agreement with that defined by the string of HH7-11.  We therefore hypothesize an association of the maser activity with continuing  outflow activity very close to the star which ultimately is responsible for the excitation of HH7-11.

	Since the VLBA imaging process results in loss of accurate position measurement, the relative positions of maser to continuum structures must be made on the basis of non-VLBA observations.  The most accurate measurement would simultaneously measure a 1.3cm counterpart of the 3.6cm source measured by Rodriguez et al. (1999) assumed to locate the star, and the maser. A ten minute observation made 1998 August 12 failed to detect the continuum emission.  Less accurately, we may determine the absolute position of the unresolved masers from the VLA observations and compare the flux-weighted centroid with that of the VLBA observations, using the offset to correct the VLBA maser positions to the more accurate VLA reference frame. For the 1998 August 12 data the maser is dominated by emission at 5.1 and 4.5 km s$^{-1}$.  One month later the strongest maser (Group I) lay at 5.3 km s$^{-1}$; we surmise that little change in maser structure occurred.  Therefore, the strongest maser lies within 0.25" of the 2.7mm emission from SVS13A1 measured by Looney, Mundy and Welch (2000) and within 0".17 of the 3.6cm emission measured by Rodriguez et al. (1999) from that source, and between these two continuum positions and within the margin of error.

	Over the fifty days of the observations the arc structure of Group II maintained a remarkably consistent morphology.  Relative to the bright reference maser in Group I, the arc consistently moved toward the ESE at a proper motion of 0.023 mas/day.  At the distance of NGC1333, or an apparent velocity in the plane of the sky of 13.6 km s$^{-1} \pm  1.7$km s$^{-1}$.  Including the small radial velocity difference between the arc material and that of the ambient material in the cloud, we estimate that the arc moved at a space velocity of 13.7 km s$^{-1} \pm  1.7$km s$^{-1}$ and that the axis of this outflow component is inclined toward us on the ESE side at an angle of $5.7^o \pm 1.6^o$.

	Toward the end of the 43m monitoring period, a feature at velocities more positive than ambient appeared.  The VLBA images show this feature slightly to the north of the others, comprised of several maser spots, and present only in the latter two epochs.  During Epoch III, two spots are seen, and during Epoch IV, five spots can be identified.  The two Epoch III spots are clearly counterparts of two of the Epoch IV spots, occurring at nearly the same position and velocity.  Over the 18 days between epochs, the V$_{LSR}$ 14.1 km s$^{-1}$ maser traveled 0.36 AU, averaging 35 km s$^{-1}$ proper motion.  The maser at 12.6 km s$^{-1}$  averaged 29 km s$^{-1}$ in proper motion.  The total space velocities of the masers were 35.3 km s$^{-1}$ and 29.4 km s$^{-1}$  at inclination of $9.5^o \pm 0.6^o$.

	Both redshifted and blueshifted masers are clearly part of the same lobe of the flow; their observed Doppler shifts are artifacts of the opening angle of the flow.  Our data suggest that the opening angle is about 10$^o$.  We measure lower expansion velocities and a wider opening angle than has been found in some other flows near Class 0 SED objects, perhaps indicating that the flows open and slow as they age.

\begin{figure}

\plotfiddle{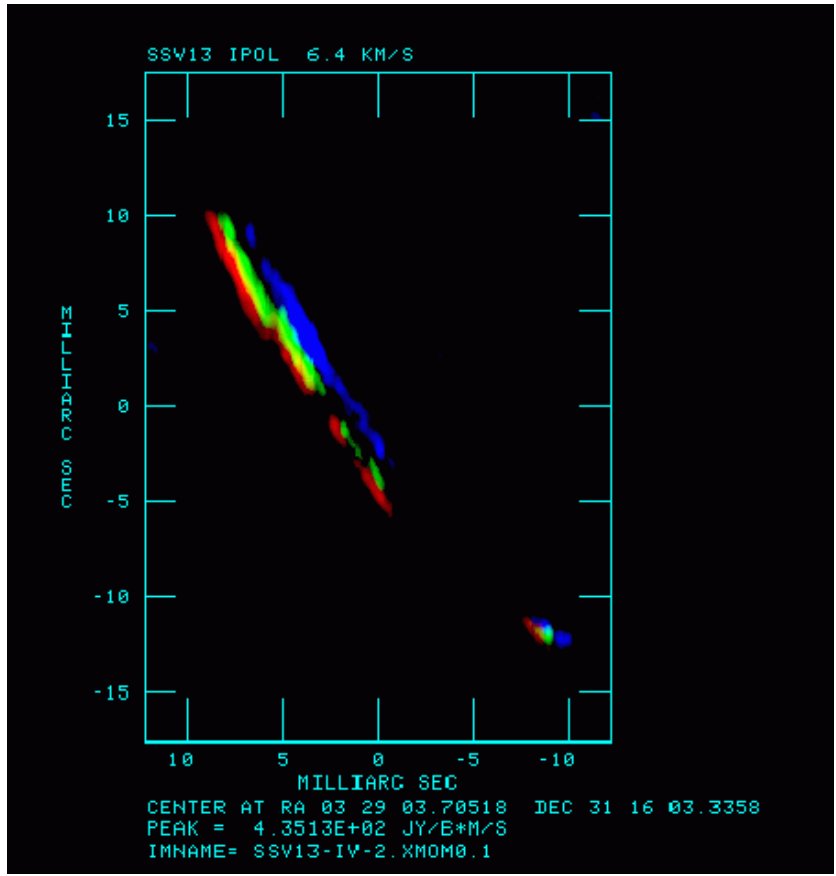}{4.0in}{0}{69}{69}{-216}{-144}
\vspace{1.0in}
\caption{VLBA images of the Group II masers at three different epochs.  Over 50 days, the arc structures consistently  moved at a rate of  0.023 mas/day, or 13.6 km/s, toward the ESE of the strong masers in Group I to the south.
Including the radial velocity offset, a space velocity of 13.7 km/s is calculated at an inclination of 6 degrees from the plane of the sky.}

\end{figure}


\begin{references}
\reference Bachiller, R., 
Guilloteau, S., Gueth, F., Tafalla, M., Dutrey, A., Codella, C., \& 
Castets, A.\ 1998, \aap, 339, L49 
\reference Codella, C., Bachiller, R., \& Reipurth, B.\ 1999, \aap, 343, 585 
\reference Haschick, A.\ D., 
Moran, J.\ M., Rodriguez, L.\ F., Burke, B.\ F., Greenfield, P., \& 
Garcia-Barreto, J.\ A.\ 1980, \apj, 237, 26 
\reference 
Jennings, R.\ E., Cameron, D.\ H.\ M., Cudlip, W., \& Hirst, C.\ J.\ 1987, 
\mnras, 226, 461 
\reference  Knee, L.\ B.\ G.\ \& 
Sandell, G.\ 2000, \aap, 361, 671 
\reference  Looney, L.\ 
W., Mundy, L.\ G., \& Welch, W.\ J.\ 2000, \apj, 529, 477 
\reference Rodriguez, L.\ F., Anglada, G., \& Curiel, S.\ 1997, \apjl, 480, L125
\reference 
Rodr{\'i}guez, L.\ F., Anglada, G., \& Curiel, S.\ 1999, \apjs, 125, 427 
\end{references}
\end{document}